\documentclass[aps,prx,twocolumn,floats,showpacs,superscriptaddress,nofootinbib]{revtex4-1}
\usepackage{graphicx,epsfig}
\usepackage{times,bbm}
\usepackage{graphics,dcolumn,bm,float}
\usepackage{amssymb,amsmath,rotate,color}
\usepackage[title,titletoc,toc]{appendix}
\usepackage{mathtools}
\usepackage{booktabs}
\usepackage{tcolorbox}
\usepackage[pagebackref=false,colorlinks,linkcolor=magenta,citecolor=blue,urlcolor=magenta]{hyperref}

\usepackage{wrapfig}
\usepackage{lipsum}
\usepackage{mwe}
\usepackage[mathlines]{lineno}
\usepackage{hyperref}
\usepackage{breqn}
\usepackage{bbold}
\usepackage{pgfplots}
\usepackage{float}
\usepackage{tkz-euclide}
\usepackage{braket}
\usepackage{physics}
\usepackage[caption=false]{subfig}
\usepackage[export]{adjustbox}
\usepackage{tikz}
\usetikzlibrary{through,calc}
\usetikzlibrary{positioning}
\usepackage[utf8]{inputenc}
\usepackage[english]{babel}
\usepackage{xcolor}

\newcommand{\be}{\begin{equation}}
\newcommand{\ee}{\end{equation}}

\newcommand{\bea}{\begin{eqnarray}}
\newcommand{\eea}{\end{eqnarray}}
\newcommand{\nn}{\nonumber}

\newcommand{\pr}{\partial}

\newcommand{\bmt}{\left[\begin{matrix}}
\newcommand{\emt}{\end{matrix}\right]}

\begin{document}
\preprint{}
\title{Holographic Hydrodynamics of {\it Tilted} Dirac Materials}
\author{A. Moradpouri}
\affiliation{Department of Physics$,$ Sharif University of  Technology$,$ Tehran$,$ P.Code 1458889694$,$ Iran}

\author{S.A. Jafari}
\email{jafari@sharif.ir}
\affiliation{Department of Physics$,$ Sharif University of  Technology$,$ Tehran$,$ P.Code 1458889694$,$ Iran}

\author{Mahdi Torabian}
\email{mahdi.torabian@sharif.ir}
\affiliation{Department of Physics$,$ Sharif University of  Technology$,$ Tehran$,$ P.Code 1458889694$,$ Iran}
\affiliation{International Centre for Theoretical Physics$,$ Strada Costiera 11$,$ 34151 Trieste$,$ Italy}

\date{\today}

\begin{abstract}

We present a gravity dual to a quantum material with tilted Dirac cone in 2+1 dimensional spacetime. In this many-body system the electronics degrees of freedom are strongly-coupled, constitute a Dirac fluid and admit an effective hydrodynamic description. The holographic techniques are applied to compute the thermodynamic variables and hydrodynamic transports of a fluid on the boundary of an asymptotically anti de Sitter spacetime with a boosted black hole in the bulk. We find that these materials exhibit deviations from the normal Dirac fluid which rely on the tilt of the Dirac cone. In particular, the shear viscosity  to entropy density ratio is reduced and the KSS bound is violated in this system. This prediction can be experimentally verified in two-dimensional quantum materials ({\it e.g.} organic $\alpha$-({BEDT}-{TTF})$_2$I$_3$ and $8Pmmn$ borophene) with tilted Dirac cone.  
\end{abstract}



\maketitle
\narrowtext

\section{Introduction}

Recent condensed matter experiments have revealed interesting electronic systems in which the low energy dynamics of quasiparticles are described by the Dirac equation. The band structure forms two cones with tips joining at the corners of the Brillouin zone in momentum space {\it a.k.a.} the Dirac points. Near band crossing the Fermi surfaces shrink to a point where electrons become effectively massless and admit a linear dispersion relation (the speed of lights is replaced by an energy-independent Fermi velocity about few hundredths of speed of light). These class of quantum materials are called Dirac materials or Dirac solids~\cite{WehlingReview}. These materials can be two-dimensional sheets that support emergent Dirac-like fermions in $2+1$ dimensional spacetime or three-dimensional materials that support Dirac/Weyl-like fermions in $3+1$ dimensional spacetime. The symmetries of the lattice at the atomic scale plays an important role. The prominent two-dimensional material is graphene ~\cite{KatsnelsonBook2012}. The density of electrons/holes in such materials can be externally tuned by changing the chemical potential away from the tip of the cone (the charge neutrality point). The chemical potential can be tuned to very small values whereby the kinetic energy (size of Fermi circle) becomes arbitrarily small and hence the ratio of potential interactions to the kinetic energy can be arbitrarily large. In fact, the Coulomb interactions are not screened effectively and the many-body system is strongly-coupled. Consequently, one expects pronounced deviations from Landau's Fermi-liquid theory of non-interacting quasi-particles as was confirmed in experiments \cite{Crossno_2016}. 

In fact, with fast moving electrons/holes, the collision rate among quasi-particles (equals inverse Planckian time) grows linearly with temperature and is much greater than the rate of momentum relaxation to phonons and impurities. It establishes a local thermodynamical equilibrium and show a collective fluid-like behavior which can be describe by relativistic hydrodynamics equations. If the Fermi temperature is smaller than the electrons temperature, we find a quantum-critical two-dimensional electronic system \cite{Lucas:2017idv}. The quasi-relativistic plasma of  Dirac particles are expected to have universal hydrodynamic transport properties. In fact, the strongly-interacting Dirac fluid is close to a perfect fluid and is expected to have a small value of shear viscosity. 

Shear viscosity is the resistivity to establishing transverse gradients in the fluid velocity and is a measure of strength of interactions among excitations in a liquid. It results in dissipations and a perfect fluid is the one with the least possible viscosity. One would expect a  theoretical lower bound on viscosity of a quantum liquid at a given temperature. On dimensional grounds, one would find for shear viscosity to entropy density ratio $\frac{\eta}{s}\sim\frac{\hbar}{k_B}$. The exact value cannot be computed by conventional perturbative field theory techniques. 

The AdS/CFT correspondence offers an alternative to studying the strongly-coupled quantum many-body systems through holographic dualities \cite{Maldacena,GKP,Witten,Witten:1998zw}. In this framework, a strongly coupled gauge theory (large color and large 't Hooft coupling) is holographically dual to a weakly coupled classical gravity. Kovtun, Son and Starinets (KSS) conjectured that there is a universal lower bound on shear viscosity to entropy ratio, {\it i.e} $\frac{\eta}{s}\geq \frac{1}{4\pi}\frac{\hbar}{k_B}$, in an infinitely strongly coupled field theory with an Einstein gravity dual \cite{KSS-1,KSS-2} (see also \cite{PSS,Buchel:2003tz}). 
The bound is respected by all known fluids in experiments. The quark-gluon plasma \cite{Shuryak:2003xe} and cold atomic Fermi gases in unitarity point \cite{,Cao:2010wa}, which are strongly coupled systems, nearly saturate the bound and are closest to perfect fluids (see \cite{Schafer:2009dj,Cremonini:2011iq} for review). There are also indications that electrons in two-dimensional metals behave as a perfect fluid \cite{Muller2009}. It would be interesting to find both theoretical models or experimental samples that could saturate or violate the bound. In fact the bound could be violated through modifications to Einstein gravity; for instance in higher-derivative gravity \cite{Kats:2007mq,Brigante:2007nu,Buchel:2003tz}, massive gravity  \cite{Alberte:2016xja} and axion/dilaton gravity \cite{Rebhan:2011vd,Jain,Jain2,Mamo,Critelli,Erdmenger}. They correspond to field theory duals with various corrections as well as with broken symmetries. 

On the other hand, it turns out that there could exist quantum materials with {\it tilted} Dirac cones ~\cite{Katayama2006,Zhou2014,Goerbig2008}. The tilt of the Dirac cone is a proxy for an emergent spacetime geometry~\cite{Volovik2016,Tohid2019Spacetime}. In fact, the low energy dynamics of electronic degrees of freedom is given by a modified Dirac Hamiltonian and the effect of tilting can be represented via substituting the Minkowski metric by a non-diagonal metric ~\cite{Tohid2019Spacetime,Jafari2019,Farajollahpour_2020}. The new metric is attributed to an emergent spacetime geometry~\cite{OjanenPRX}.  

In this paper we study the hydrodynamic limit of quantum-critical electronic systems with tilted Dirac cones. This many-body system is strongly-coupled and we expect to see a fluid-like behavior. We propose a holographic gravity dual of the system. We find interesting modifications in thermodynamical variables and hydrodynamical transports compared to non-tilt materials. In fact, we find that the shear viscosity to entropy density ratio is reduced compared and the KSS bound is violated in tilt fluids. It is known that the smaller the viscosity, the higher the tendency to turbulent current flow. In principle, this effect could be investigated in future experiments. We see that  computationally impenetrable problems of strongly-coupled electronic systems can be approached using string theory techniques. In return, condensed matter experimentations set a valuable stage to verifying string theory predictions. 

The structure of the paper is as follows; In the next section we introduce an AdS-Schwarzschild black-hole geometry as the gravity dual to the tilted Dirac materials at finite temperature. We compute temperature, entropy, energy-momentum tensor and the shear viscosity of tilted Dirac material through the holographic prescriptions. Finally, we summarize in the last section with prospects for future experiments.

\section{A Holographic model}
In this section we propose a holographic model for 2+1 dimensional quantum materials tilted cones. 
These materials are parameterized by two-parameter family of tilt deformations
$(\zeta_x,\zeta_y)$. At the vicinity of the Dirac points the degrees of freedom are semi-relativistic and are described by a deformed Dirac Hamiltonian. It is found that these systems admit a covariant spacetime description with a non-trivial metric such that the tilt effect can be implemented through using the following metric tensor (and its inverse) \begin{equation}
\label{metric2.eqn}
g_{\mu\nu}=
\begin{pmatrix}
-1+\zeta^2 & -\zeta_j\\
-\zeta_i & \delta_{ij}
\end{pmatrix},
g^{\mu\nu}=
\begin{pmatrix}
-1 & -\zeta_i\\
-\zeta_j & \delta_{ij}-\zeta_i\zeta_j,
\end{pmatrix},
\end{equation} 
in equations involving the Minkowski metric $\eta_{\mu\nu}$  (and its inverse), where $\zeta^2=\zeta_x^2+\zeta_y^2$ and $i=x,y$.  At long wavelength and short frequency the electronic system behaves like a fluid of Dirac particles (see \cite{Lucas:2017idv,Moradpouri:2020wwa}). This many-body system is strongly-coupled. In the holographic framework of AdS/CFT correspondence, it is conjectures that there exist a classical weakly-coupled gravity dual to such a system \cite{Maldacena,GKP,Witten,Witten:1998zw}. In this setup, the three-dimensional field theory at finite-temperature lives on the boundary of a four-dimensional spacetime with is asymptotically AdS which has a black hole in bulk. 

The Einstein gravity with a negative cosmological constant $-6/L^2$ in four dimensions has an $AdS_4$ solution with with curvature radius $L$. There exist also solutions with black holes in the interior which are asymptotically AdS.  These gravitational backgrounds represent the gravity duals of thermal field theories. There is a three-parameter family of Schwarzschild-AdS geometries in four dimensions 
which are parametrized by the black-hole temperature and two boost factors along two spatial dimensions of the boundary \cite{Hubeny}.
These are precisely the hydrodynamical variables about thermal equilibrium a fluid, namely $T,u^\mu$ (with a normalization constraint $u^\mu u_\mu=-1$) in terms of which the fluid energy-momentum tensor is written. In the other words, the metric solution induces the boundary energy-momentum tensor. The solutions are given by
\begin{equation}
ds^2=\frac{1}{L^2z^2}\Big[-f(z)u_{\mu}u_{\nu}+{\cal P}_{\mu\nu}\Big]{\rm d}x^{\mu}{\rm d}x^{\nu}+\frac{L^2}{z^2f(z)}{\rm d}z^2,
\end{equation}
where ${\cal P}^{\mu\nu}=\eta^{\mu\nu}+u^{\mu}u^{\nu}$ and the blackening factor $f(z)=1-(z/z_h)^3$. The AdS boundary is locates at $z=0$ and the black hole horizon at $z=z_h$.

To obtain the gravity dual of a tilted Dirac material we replace the Minkowski metric $\eta_{\mu\nu}$ with metric $g_{\mu\nu}$ given in \eqref{metric2.eqn}. This insight is based on previous results that the tilted Dirac system admit an emergent spacetime description which is endowed with metric $g_{\mu\nu}$ \eqref{metric2.eqn}. We take the black hole at rest such that $u^{\mu}=\frac{1}{\sqrt{1-\zeta^2}}(1,\vec 0)$ and, for the sake of computational simplicity, we assume $\zeta_y=0$ (thus $\zeta=\zeta_x$) to find
\bea\label{metric} {\rm d}s^2 &=& \frac{L^2}{z^2}\frac{{\rm d}z^2}{f(z)}+\frac{1}{z^2L^2}\Big[ - (1-\zeta^2)f(z){\rm d}t^2 - 2\zeta f(z) {\rm d}t{\rm d}x\cr
&&\qquad\qquad\qquad\qquad +\frac{1-\zeta^2f(z)}{1-\zeta^2}{\rm d}x^2+{\rm d}y^2\Big].\eea
This metric solution sets the background above which we can systematically study long scale fluctuations and responces in the rest of the section.

\subsection{Holographic Thermodynamics}
In gauge/gravity duality, tinite temperature field theory is dual to black hole thermodynamics in asymptotically AdS spacetime. The Hawking temperature and the Bekenstein-Hawking entropy of a black-hole are identified respectively with the temperature and entropy in the field theory. 
To read the thermodynamic quantities, we change the signature of the metric by introducing the Euclidean time $\tau=it$ and simultaneously $\tilde\zeta= i\zeta$ using the singularity approach~\cite{Hawking}. The relevant part of the metric reads as
\bea {\rm d}s^2 &\supset& \frac{L^2}{z^2}\frac{{\rm d}z^2}{f(z)} +\frac{1}{z^2L^2}\Big[ (1+\tilde\zeta^2)f(z){\rm d}\tau^2 \!+\! 2\tilde\zeta f(z){\rm d}\tau{\rm d}x\Big].\quad\eea
To avoid singularity at the horizon, $\tau$ must be periodic $\tau\sim\tau+\beta$ where $\beta^{-1} =k_BT$. We introduce a new coordinate r which measures distance from the horizon and a canonical angular coordinate $\theta$  as
\bea {\rm r} = \frac{2L}{\sqrt3}\sqrt{1-\frac{z}{z_h}},\qquad \theta = \frac{3\sqrt{1+\tilde\zeta^2}}{2L^2z_h}\tau.\eea
The near horizon geometry is given in polar coordinates ${\rm d}s^2 = {\rm dr}^2 +{\rm r}^2 {\rm d}\theta^2+\cdots$ with $\theta\sim\theta+2\pi$.
Then, 
we find the temperature as
\bea k_BT = \frac{1}{L^2z_h}\frac{3}{4\pi}\sqrt{1-\zeta^2}.\eea
We observed a very pecular dependence of the field theory temperature on the tilt parameters. 

On the other hand, the Bekenstein-Hawking entropy is given by the horizon area
\bea\label{entropy-1} S_{\rm BH} \!=\! \frac{k_B}{\hbar G_N}\frac{1}{4}\! \int\!{\rm d}^2x\sqrt{g(z_h)} =\!\frac{k_B}{\hbar G_N}\frac{1}{4L^2z_h^2\sqrt{1-\zeta^2}}V_2,\quad\ \eea
where $V_2$ is a two-dimensional area. Then, the entropy density is $s=S_{\rm BH}/V_2$. Again, we find a special dependence of the entropy density on the tilt parameters which grows singular as the tilt approaches to one.  

We can also compute the field theory partition function from the on-shell gravity action in Euclidean signature a al holographic GKP-W relation \cite{GKP,Witten} 
\bea {\cal Z}[\varphi_b] = \exp(-{\cal S}^{\rm on-shell}_{E}[\varphi(z=0)=\varphi_b]/\hbar).\eea 
As we integrate over infinite volume of the bulk the on-shell action is divergent. We regularize the action by introducing a cut-off at $z=\epsilon$ and the divergences are renormalized away by adding counter-terms. In fact, we stay an $\epsilon$ distance from the boundary, compute the bulk integral up to $z\geq\epsilon$ and the boundary terms at $z=\epsilon$ and finally take $\epsilon$ to zero. 

The regularized gravitational action in a four dimensional manifold with a boundary includes the Einstein-Hilbert action, the Gibbons-Hawking boundary term and the counter-terms ~\cite{deHaro,Balasubramanian,ammon2015gauge}
\bea\label{action} {\cal S}_E &\!=\!&- \frac{1}{16\pi G_N}\!\int_0^\beta\!\!{\rm d}\tau\!\int_{z=\epsilon}\!\!{\rm d}{\bf x}\bigg[\int_{z_h}^\epsilon\!{\rm d} z\sqrt{-g}\Big(\!R+\!\frac{6}{L^2}\Big)
\cr &&\qquad\qquad\qquad\qquad -2\sqrt{-\gamma}K\!+\!2 \sqrt{-\gamma}\Big(\frac{4}{L}+L R\Big)\!\bigg].\quad\ \ \eea
The boundary of AdS is now located at $z=\epsilon$, the induced metric on this boundary is $\gamma_{ij}=\frac{L^2}{\epsilon}g_{ij}(\epsilon,x)$ and $K=\gamma^{ij}\nabla_i n_j$ is the trace of its extrinsic curvature. The non-vanishing component of the unit outgoing normal vector at the boundary is $n^z=L^{-1}z\sqrt{f(z)}$. Then, via the method of holographic renormalization we find 
\bea {\cal S}_E^{\rm on-shell} &\!=& \!\lim_{\epsilon\rightarrow 0}({\cal S}_{\rm EH} \!+\! {\cal S}_{\rm GH} \!+\! {\cal S}_{\rm c.t.})\!= \!-\frac{V_2}{16\pi G_Nk_BTz_h^3L^4}.\quad\ \eea
It is basically the Helmholtz free-energy $F=-k_BT\ln{\cal Z} = k_B{\cal S}_E^{\rm on-shell}/\hbar$ in natural units. The entropy is computed through $S=-(\partial F/\partial T)_V$ as
\bea\label{entropy-2} S = - \frac{k_B}{\hbar}\frac{\partial}{\partial T}\Big(T{\cal S}_E^{\rm on-shell}\Big)=\frac{k_B}{\hbar G_N}\frac{1}{4L^2z_h^2\sqrt{1-\zeta^2}}V_2,\ \ \eea
which gives \eqref{entropy-1}. The entropy density is then $s=S/V_2$.

The energy density $\epsilon = U/V_2=(F+TS)/V_2$ is
\bea \varepsilon \!=\! \frac{k_BT}{\hbar}\Big({\cal S}_E^{\rm on-shell} - \frac{\partial(T{\cal S}_E^{\rm on-shell})}{\partial T}\Big)\frac{1}{V_2}
\!=\! \frac{k_B}{8\pi G_N\hbar z_h^3L^4},\quad\ \ \eea
and the pressure density $p=-(\partial F/\partial V)_{T}$ is computed as
\bea p = -\frac{k_B}{\hbar}\frac{\partial (T{\cal S}_E^{\rm on-shell}) }{\partial V} = \frac{k_B}{16\pi G_N\hbar z_h^3L^4}.\eea
We find that $\varepsilon=2p$ via the holographic approach. Moreover, we see that the tilted Dirac material satisfies the first law of thermodynamics $\varepsilon+p=Ts$.

The full boundary energy-momentum tensor can be read from functional derivative of the on-shell gravity action in the presence of the metric $g_{(0)}^{ij}$ as a source
\bea \langle T_{\mu\nu}\rangle\! &=&\! 
 -\frac{2}{\sqrt{g_{(0)}}}\frac{\delta{\cal S}_{\rm E}}{\delta g^{\mu\nu}_{(0)}}\! =
\lim_{\epsilon\rightarrow0}\frac{L}{\epsilon^{1/2}}T^{\gamma}_{\mu\nu}(\epsilon) = \frac{3L^2}{16\pi G_N} g^{(3)}_{\mu\nu},\qquad\eea
where $T_{ij}^{\gamma}$ is the energy-momentum tensor at $\epsilon$ computed with the induced metric $\gamma$. The metric $g^{(3)}_{\mu\nu}$ is read from the near boundary expansion of the the asymptotically AdS metric
\bea {\rm d}s^2 &=& \frac{L^2}{\tilde z^2}\Big[{\rm d}\tilde z^2+g_{\mu\nu}(x,\tilde z){\rm d}x^\mu{\rm d}x^\nu\Big],\\
 g_{\mu\nu}(x,\tilde z) &=& g^{(0)}_{\mu\nu}(x) \!+\! \tilde zg^{(1)}_{\mu\nu}(x) \!+\! \tilde z^2g^{(2)}_{\mu\nu}(x)\!+\!\tilde z^3g^{(3)}_{\mu\nu}(x)\!+\!\cdots,\nn\eea
which is regular as $\tilde z\rightarrow0$. 
We can rewrite the metric \eqref{metric} in the above form through a change of coordinate as
\bea z = \frac{\tilde z}{L^2\big(1+\frac{\tilde z^3}{4L^6z^3_h}\big)^{2/3}},\eea
such that near the boundary and keeping leading orders in $\tilde z$ gives
\bea {\rm d}s^2 \!&=&\! \frac{L^2}{\tilde z^2}\bigg[{\rm d}\tilde z^2 - (1-\zeta^2)\Big(1-\frac{2\tilde z^3}{3z_h^3}\Big){\rm d}t^2 -
 2\zeta \Big(1-\frac{2\tilde z^3}{3z_h^3}\Big) {\rm d}t{\rm d}x \cr &&\quad +\Big(1+\frac{(1+2\zeta^2)}{3(1-\zeta^2)z_h^3}\tilde z^3\Big){\rm d}x^2+\Big(1+\frac{\tilde z^3}{3z_h^3}\Big){\rm d}y^2\bigg].\nn\\\eea 
Then, the non-zero components of the energy momentum tensor are
\bea
T_{tt}&&=\frac{(1-\zeta^2)}{8\pi G_N z^3_hL^4},~~~T_{0x}=\frac{\zeta }{8\pi G_N z^3_hL^4}\nn\\
T_{xx}&&=\frac{(1+2\zeta^2)}{16\pi G_N(1-\zeta^2)z^3_hL^4},~~T_{yy}=\frac{1}{16\pi G_N z^3_hL^4}.
\eea
It can be easily seen that the energy-momentum is traceless $T_{\mu}^{\mu}=0$. For the perfect fluid at rest, $u^{\mu}=\frac{1}{\sqrt{1-\zeta^2}}(1,\vec 0)$, in the tilted Dirac background, the energy density and pressure is given by
\bea
T_{tt}=(1-\zeta^2)\varepsilon,~~~T_{yy}=p,
\eea
so we again find the equation of state $\varepsilon=2p$.

\subsection{Holographic Hydrodynamics}
In this part we study the hydrodynamics of tilted Dirac material using holographic dualities. In particular, among transport coefficients, we compute the shear viscosity of the fluid. 

\subsubsection{Shear viscosity, linear response and the Kubo relation}
The energy-momentum tensor of a viscous fluid in the Landau frame is 
\bea\label{EMT} T^{\mu\nu} = \varepsilon u^\mu u^\nu +p{\cal P}^{\mu\nu} + \sigma^{\mu\nu},\eea
where ${\cal P}^{\mu\nu}=g^{\mu\nu}+u^\mu u^\nu$ is the projection tensor and $u^\mu u_\mu=-1$. The dissipation tensor $\sigma^{\mu\nu}$ includes the shear viscosity tensor $\eta^{\mu\nu\rho\sigma}$ which to the leading order in derivative expansion can be reads from a general dissipation function
\bea
	F=\frac{1}{2}\eta^{\mu\nu\rho\sigma}u_{\mu\nu}u_{\rho\sigma},
\eea
where $u_{\mu\nu}=\frac{1}{2}(\nabla_{\mu}u_{\nu}+\nabla_{\nu}u_{\mu})$
and $\nabla$ is the covariant derivative with respect to metric $g_{\mu\nu}$.
Then,
\bea\label{sigma}
	\sigma^{\mu\nu}=-\frac{\partial F}{\partial u_{\mu\nu}}=-\eta^{\mu\nu\rho\sigma}u_{\rho\sigma}.
\eea

 In the Landau frame we have $\sigma_{\mu\nu}u^\mu=0$ which in the fluid rest frame
 implies $\sigma_{00}=\sigma_{0i}=0$. Thus, all dissipative effects are in $\sigma_{ij}$ which in a system with two spatial dimension would only be $\sigma_{xy}$. 
The shear viscosity, as a phenomenological transport coefficient, parametrizes the response of a system in thermal equilibrium to infinitesimal perturbations by external source. Thus, to compute shear viscosity we apply homogeneous time-dependent metric fluctuations $h_{xy}(t)$ and study the response of $T_{xy}$ in a fluid at rest. We must be cautious that, whereas there is an intrinsic anisotropy and explicit violation of parity invariance, the other hydrodynamical parameters including $\delta p(t)$, $\delta \varepsilon(t)$(=2$\delta p(t)$) and $\delta v^i(t)$ may be perturbed too.
Then, the corresponding perturbed components of energy-momentum tensor are 
\bea
\delta T^{00}&&=\frac{2+\zeta^2}{1-\zeta^2}\delta p-\frac{6p\zeta}{(1-\zeta^2)^{\frac{3}{2}}}\delta v^x,\\
\delta T^{0x}&&=\frac{3p}{\sqrt{1-\zeta^2}}\delta v^x-\zeta\delta p,\\
\delta T^{0y}&&=\frac{3p}{\sqrt{1-\zeta^2}}\delta v^y+\zeta p h_{xy},
\eea
where we have used $\delta u^{\mu}=(-\frac{1}{1-\zeta^2}\vec\zeta.\delta\vec v, \delta\vec v^i)$ such that normalizability condition $g_{\mu\nu}u^{\mu}u^{\nu}=-1$ is satisfied.
The conservation of energy-momentum for homogeneous perturbations, $\nabla_{\nu} \delta T^{\nu\mu}=0$, implies
\bea
\label{energy-momentumequations}
\partial_0[\delta T^{00},\delta T^{0x}] &=&0,\\
\pr_0\delta T^{0y}\!+\!\Big(\Gamma^t_{ty}\!+\!\Gamma^x_{xy})T^{yy}\!+\!2\Gamma^y_{tx}T^{tx}\Big)&=&0.\quad\ 
\eea

As the non-zero components of Christoffel connection (assuming $\zeta_y=0$) are
\bea
\Gamma^{x}_{xy}&=&-\Gamma^{t}_{ty}=\frac{\zeta}{2}\pr_{t}h_{xy}, \cr \Gamma^{t}_{xy} &=& \Gamma^{y}_{xt}=\frac{1}{2}\pr_{t}h_{xy},\\ \Gamma^{x}_{ty}&=&\frac{1}{2}(1-\zeta^2)\pr_{t}h_{xy},\nn\eea
then, the conservation equations imply 
\bea\partial_0\delta p=\partial_0\delta v^x=\partial_0\delta v^y=0.\eea
It implies that the energy and pressure density are unperturbed and the fluid remain at rest. Then, we compute the response $T_{xy}$ to perturbation $h_{xy}$ to the leading order using \eqref{EMT} and \eqref{sigma} as (note that $h^{xy}=-(1-\zeta^2)h_{xy}$)
\bea\label{deltaT} \delta T^{xy}=-p(1-\zeta^2)h_{xy}-\frac{\eta^{xyxy}}{\sqrt{1-\zeta^2}}\partial_th_{xy}.\eea
In the Fourier space it implies
\bea \delta \sigma^{xy}(\omega) = i\omega \eta^{xyxy}(\omega) h_{xy}(\omega),\eea
where we dropped the delta function term from the first term of \eqref{deltaT}.  In the rest of the paper we define $\eta=\eta^{xyxy}$.

On the other hand, a la linear response theorem, each transport coefficient is related to a retarded Green's function. For homogeneous perturbations, namely zero-momentum-transfer, linear response implies (in the Fourier space)
\bea\label{liners-response} \delta T_{xy} (\omega) = G^R_{T_{xy},T_{xy}}(\omega)h_{xy}(\omega),\eea
where the two-point function is
\bea G_{T_{xy},T_{xy}}^R\!(\omega) \!=\! -iV_2\!\!\int\!\! {\rm d}t e^{i\omega t}\theta(t) \langle [T_{xy}(t),T_{xy}(0)]\rangle.\ \eea
Then, we compare equations \eqref{deltaT} and \eqref{liners-response} to find the so called Kubo formula
\bea\label{eta} \eta = -\sqrt{1-\zeta^2}\lim_{\omega\rightarrow 0}\frac{1}{\omega}{\rm Im\,}G_{T_{xy}T_{xy}}^R(\omega).\eea
In the next section, we follow the holographic approach to computing the retarded Green's function in a dual gravitational theory.

\subsubsection{Holographic retarded Green's function and reduced  $\eta/s$}
Here we study the propagation of gravitational waves in an asymptotically AdS spacetime with a Schwarzschild black hole in the interior. The metric perturbation that helped to study the response in energy-momentum tensor in the previous section has an interpretation in the holographic framework; perturbations in boundary energy-momentum tensor is dual to gravitational waves in the bulk. The holographic renormalization helps to compute the one-point function in the presence of a source.  The leading and sub-leading terms in the near boundary expansion correspond respectively to the external source and the response. 
The retarded Green's functions are found by imposing in-falling boundary condition at the horizon. 

We consider time-dependent homogeneous metric perturbation $h_{xy}(t,{\bf x},u)=h_{xy}(t,u)$ which propagates in the bulk and deform the metic as
\begin{equation}
{\rm d}s^2={\rm d}s_{\rm AdS-BH}^2+2h_{xy}(t,u){\rm d}x{\rm d}y.
\end{equation}
We define a new coordinate $u=z/z_h$ such that $u=0$ at the AdS boundary and $u=1$ at the black hole horizon. We apply a Fourier transformation along temporal direction and write the perturbations as
\bea 	\label{Fourier1.eqn}
h_{xy}(t,u)=\frac{1}{L^2u^2}\int \frac{{\rm d}\omega}{2\pi}\phi_\omega(u)e^{-i\omega t}.
\eea
Due to non-diagonal form of the metric on the boundary, the $ty$-component of the Einstein equation is not satisfied as $G_{ty}\sim \zeta h_w\neq0$. We can alleviate the problem if we consider $h_{ty}$ perturbation as well (we recall that for the sake of simplicity we have set $\zeta_y=0$)
\begin{equation} 
{\rm d}s^2={\rm d}s_{\rm AdS-BH}^2+2[h_{xy}(t,u){\rm d}x+h_{ty}(t,u){\rm d}t]{\rm d}y.
\end{equation}
Similarly, we apply a partial Fourier transformation such that
\bea	\label{Fourier2.eqn}
h_{ty}(t,u)=\frac{1}{L^2u^2}\int \frac{{\rm d}\omega}{2\pi}\psi_\omega(u)e^{-i\omega t}.\eea
To be consistent with the dual field theory picture, this perturbation is vanishing on the boundary at $u=0$.
Then, the Einstein equations imply
 \bea
&&\psi'_\omega=\zeta\frac{(1-u^3)(1-u^3)}{1-(1-u^3)\zeta^2}\phi'_\omega\label{G23},\\
&&\psi''_\omega-\frac{2}{u}\psi'_\omega+\zeta\frac{L^4\omega^2}{1-u^3}\phi_\omega=0\label{G02},\\
&&\phi''_\omega-\frac{2+u^3}{u(1-u^3)}\phi'_\omega+\frac{L^4\omega^2(1-(1-u^3)\zeta^2)}{(1-u^3)^2(1-\zeta^2)}\phi_\omega\label{G12}\\
&&\qquad\qquad\qquad\qquad\qquad\qquad\qquad=\zeta\frac{3 u^2}{(1-u^3)(1-\zeta^2)}\psi'_\omega,\nn
\eea
where $'$ is the derivative with respect to $u$. The above equations are consistent and Eqs.~\eqref{G23} and~\eqref{G02} imply Eq.~\eqref{G12}.
We substitution Eq.~\eqref{G23} into the Eq.~\eqref{G02} to find  
 \bea
\phi''_\omega&-&\frac{2+u^3-2\zeta^2(1-u^3)^2}{u(1-u^3)(1-\zeta^2(1-u^3))}\phi'_\omega\label{EOM}\\
&+&\frac{L^4\omega^2(1-\zeta^2(1-u^3))}{(1-u^3)^2(1-\zeta^2)}\phi_\omega=0.\nn
\eea
We solve Eq.~\eqref{EOM} near the boundary as $u\to 0$. The solution in $\omega\to 0$ limit
, up to the first order in $\omega$, is 
\bea\phi_\omega(u)=(c_1+d_1\omega)\!+\!(c_2+d_2\omega)(\log(1-u^3)+\zeta^2u^3).\qquad 
\eea
We can use a boundary condition at the boundary $u=0$ such that $\phi_\omega(u)=\phi^{(0)}_{\omega}$ that fixes $c_1+d_1\omega$. Then, the solution is
\bea
\label{omegaexpansion}
\phi_\omega(u)=\phi^{(0)}_{\omega}\big[1+\phi^{(1)}_{\omega}(\log(1-u^3)+\zeta^2u^3)\big],
\eea
The coefficient $\phi^{(1)}_{\omega}$ cannot be fixed using the near boundary limit information, so we have to solve the Eq.~\eqref{EOM} in the near horizon limit.

In the near horizon limit $u\to 1$ Eq.~\eqref{EOM} is simplified as follows 
\bea
\phi''_\omega-\frac{1}{1-u}\phi'_\omega+\frac{L^4\omega^2}{9(1-u)^2(1-\zeta^2)}\phi_\omega=0,
\eea
For the ansatz $\phi_\omega\sim (1-u)^m$ we find
\bea
m=\pm i\frac{L^2\omega}{3\sqrt{1-\zeta^2}},
\eea
We impose incoming/infalling boundary condition at the horizon by choosing the negative sign. Thus, the solution close to the horizon is 
\bea
\label{nearhorizon}
\phi_\omega(u)&&\sim1-i\frac{L^2}{3\sqrt{1-\zeta^2}}\omega\log(1-u^3).\ 
\eea
The above solution is valid when $1-u\ll 1$ and $\log(1-u)\gg 1$. In the limit $\omega\log(1-u)\ll 1$, by comparing Eqs.~\eqref{omegaexpansion},\eqref{nearhorizon}, the near boundary solution is given by
 \bea
 \phi_\omega=\phi^{(0)}_{\omega}\Big(1+i\frac{L^2}{3}\omega\sqrt{1-\zeta^2}u^3\Big)
 \equiv \phi^{(0)}_{\omega}f_{\omega}(u).
 \eea
 Given the above solution, we use Eq.~\eqref{G23} to find $\psi_\omega(u)$ as
  \bea
 \label{sfunction}
\psi_\omega=\psi^{(0)}_{\omega}\!+\!i\frac{L^2}{3}\zeta\sqrt{1-\zeta^2}\omega\phi^{(0)}_{\omega}u^3\!\equiv \psi^{(0)}_{\omega}\! +\!\phi^{(0)}_{\omega}g_{\omega}(u).\quad\ \ 
\eea 

Using the solutions to the equations of motion, we can compute the regularized on-shell gravity action on the boundary. 
The GKP-W relation \cite{GKP,Witten}, in Euclidian signature, gives the one-point function
\bea \langle \sigma^{xy}\rangle = \lim_{\epsilon\rightarrow0}\frac{\partial{\cal S^{\rm on-shell}_{\rm reg}}}{\partial \phi^{(0)}_{xy}}.\eea
In the Lorentzian signature, we apply a related prescription \cite{Son_2002}. We bring the on-shell gravity action into the following form 
\bea {\cal S}^{\rm on-shell}_{\rm reg} = V_2\int\frac{{\rm d}\omega}{2\pi}\phi^{(0)}_\omega{\cal F}(\omega,u)\Big|_{u=\epsilon}^{u=1}\phi^{(0)}_\omega,\eea
and then according to the recipe \cite{Son_2002}, the retarded Green's function is
\bea\label{prescription} G_R(\omega) = -2\lim_{\epsilon\rightarrow 0}{\cal F}(\omega,\epsilon),\eea
In the following, we compute the action terms up quadratic terms in perturbations. Here we use the action \eqref{action} this time in the Lorentzian signature. 
\paragraph{The Einstein-Hilbert action}by expanding action  up to the second order, we have  (momentarily $8\pi G_N=1$)
\bea
	\label{secondorderaction.eqn}
	{\cal S}_{\rm EH}^{(2)}&&=\frac{2V_3}{L^4}\Big(1-\frac{1}{\epsilon^3}\Big)\cr &&+ \frac{V_2}{L^4}\!\int\!\frac{{\rm d}\omega}{2\pi}\Big[\Big(\frac{1}{\epsilon^3}-1\Big)(1-\zeta^2)\phi_\omega\phi_{-\omega}-\frac{3(1\!-\!\zeta^2)}{2\epsilon^2}\phi_{-\omega}\phi'_\omega\nn\\
	&&+\frac{3}{2\epsilon^2}\psi_{-\omega}\psi'_\omega+\Big(\frac{3-2\zeta^2}{2(1-\zeta^2)}-\frac{1}{\epsilon^3}\Big)\psi_\omega\psi_\omega\cr&&-\frac{3\zeta}{2\epsilon^2}\psi_{-\omega}\phi'_\omega-\frac{3\zeta}{2\epsilon^2}\phi_\omega\psi'_\omega+2\zeta\Big(\frac{1}{\epsilon^3}-1\Big)\psi_{-\omega}\phi_\omega\Big].\quad\
	\eea
We have regularized the on-shell action via a non-zero $\epsilon$.

\paragraph{The Gibbons-Hawking term}
The boundary term up to second order on perturbations is computed as
\bea
 S_{\rm GH}^{(2)}&&=-\frac{3V_3}{L^4}\Big(1-\frac{2}{\epsilon^3}\Big)\cr &&+\frac{V_2}{L^4}\int\frac{{\rm d}\omega}{2\pi}\Big[3\Big(\frac{1}{2}-\frac{1}{\epsilon^3}\Big)(1-\zeta^2)\phi_\omega\phi_{-\omega}\nn\\
 &&+\frac{2}{\epsilon^2}(1-\zeta^2)\phi_{-\omega}\phi'_\omega+\frac{2\zeta}{\epsilon^2}(\phi_{-\omega}\psi_\omega)'-\frac{2}{\epsilon^2}\psi_{-\omega}\psi'_\omega\nn\\
 &&-3\Big(\frac{2}{\epsilon^3}-1\Big)\zeta \phi_{-\omega}\psi_\omega+3\Big(\frac{1}{\epsilon^3}-\frac{1}{2}\Big)\psi_{-\omega}\psi_\omega\Big].
\eea

\paragraph{The counter terms}
The counter terms up to quadratic terms are found as
\bea
{\mathcal S}_{\rm c.t.}^{(2)}&&=\frac{2V_3}{L^4}\Big(1-\frac{2}{\epsilon^3}\Big)\cr &&-\frac{4V_2}{L^4}\int\frac{{\rm d}\omega}{2\pi}\Big[\Big(\frac{1}{4}-\frac{1}{2\epsilon^3}\Big)(1-\zeta^2)\phi_{-\omega}\phi_\omega\nn\\
&&+\zeta\Big(\frac{1}{2}-\frac{1}{\epsilon^3}\Big)\phi_{-\omega}\psi_\omega+\Big(\frac{1}{2\epsilon^3}\!+\!\frac{1+\zeta^2}{2(1-\zeta^2)}\Big)\psi_{-\omega}\psi_\omega\Big].\quad\ \ \eea

The full regularized  on-shell gravity action is found by adding above three contributions
\bea
\label{onshellaction}
{\cal S}^{\rm on-shell}_{\rm reg}&&={\cal S}_{\rm EH}^{(2)}+{\cal S}_{\rm GH}^{(2)}+{\cal S}_{\rm ct}^{(2)}=\frac{V_3}{L^4} 
\cr &&+\frac{V_2}{L^4}\!\int\!\frac{{\rm d}\omega}{2\pi}\Big[\!\!-\!\frac{1-\zeta^2}{2}\phi_{-\omega}\phi_\omega+\frac{1-\zeta^2}{2\epsilon^2}\phi_{-\omega}\phi'_\omega
\cr &&\qquad\quad   -\frac{2+\zeta^2}{2(1-\zeta^2)}\psi_{-\omega}\psi_\omega -\frac{1}{2\epsilon^2}\psi_{-\omega}\psi'_\omega
\cr &&\qquad\quad  +\frac{\zeta}{2\epsilon^2}\psi_{-\omega}\phi'_\omega +\frac{\zeta}{2\epsilon^2}\phi_{-\omega}\psi'_\omega -\zeta \phi_{-\omega}\psi_\omega\Big].\quad\ \ 
\eea
To be consistent with Eq.~\eqref{deltaT}, the boundary value of $\psi_\omega$ is set to zero in Eq.~\eqref{sfunction}.Then, the on shell action~\eqref{onshellaction} is simplified in terms of $\phi^{(0)}_{\omega}$ as
\bea
{\cal S}^{\rm on-shell}_{\rm reg}=\frac{V_3}{L^4}+\frac{V_2}{L^4}\int\frac{d\omega}{2\pi}\phi^{(0)}_{-\omega}\mathcal{F}({\omega},\epsilon)\phi^{(0)}_{\omega},
\eea
where 
\bea
\mathcal{F}({\omega},\epsilon)&&=-\frac{1-\zeta^2}{2}f_{-\omega}f_{\omega}+\frac{1-\zeta^2}{2\epsilon^2}f_{-\omega}f'_{\omega}\nn\\
&&-\frac{2+\zeta^2}{2(1-\zeta^2)}g_{-\omega}g_{\omega}-\frac{1}{2\epsilon^2}g_{-\omega}g'_{\omega}\nn\\
&&+\frac{\zeta}{2\epsilon^2}g_{-\omega}f'_{\omega}+\frac{\zeta}{2\epsilon^2}f_{-\omega}g'_{\omega}-\zeta f_{-\omega}g_{\omega},
\eea
Then, the Lorentzian prescription to computing the one-point function \eqref{prescription} (after restoring $8\pi G_N$ and $z_h$) implies 
\bea G^R_{T_{xy}T_{xy}} \!=\! \frac{1}{16\pi G_4L^4z^3_h}\Big[(1-\zeta^2)-i\omega L^2z_h\sqrt{1-\zeta^2}\Big].\quad\ \eea
Then, the shear viscosity is read from \eqref{eta} as
\bea\label{eta-2} \eta = \frac{1-\zeta^2}{16\pi G_4L^2z^2_h}.\eea
Finally, using above equation and either \eqref{entropy-1} or \eqref{entropy-2} the shear to entropy density ration for the tilted Dirac material is computed 
\bea 
\frac{\eta}{s}=\frac{\hbar}{k_B}\frac{(1-\zeta^2)^{\frac{3}{2}}}{4\pi}.
\eea
Apparently, the Dirac fluid of electrons/holes in a quantum matter with tilted Dirac cones has a reduced $\eta/s$ compared to the KSS bound. It is a pronounced violation of a universal bound in perfect fluids. Interestingly, it approaches zero by double scaling (increasing entropy and decreasing shear viscosity)  as the tilts goes to one. Among other things, it could result in a turbulent flow of quasi-particles. It is a major prediction for the tilted Dirac systems in a holographic framework. It can be verified in sophisticated condensed matter experiments studying two dimensional materials. 

\section{Conclusion}
In this paper we presented a holographic gravity dual to a strongly-coupled electronic system with tilted Dirac cones. Applying the holographic techniques, we computed the free energy, energy-momentum tensor and the shear viscosity of the effective Dirac fluid. We find abnormal behavior in thermodynamic variables and hydrodynamic transports for the tilted Dirac materials. They are sensitive to the tilt parameter of the quantum system. 

In principle, deviations from the normal Dirac materials can be examined in condensed matter playgrounds especially that the low sensitivity in the observables can be compensated by high values of the tilt parameters.
Candidate quantum materials with two-dimensional tilted Dirac cone to verify our prediction are 
the organic conductor $\alpha$-({BEDT}-{TTF})$_2$I$_3$ salt~\cite{Katayama2006} which is experimentally available and 8Pmmn borophene~\cite{Zhou2014} 
which is predicted to have tilted Dirac cone. It is possible to tune the chemical potential of the organic conductor to an appropriate range and dopp it in the Dirac fluid regime to get a feasible platform to checking the violation of the KSS bound. 

It is extremely interesting that techniques from high energy physics offers a solution to an impenetrable problem of a strongly-coupled condensed matter system and in return the latter offers a playground to test the predictions of the former. 

 \paragraph*{Acknowledgements} MT is thankful to the HECAP section of ICTP for warm hospitality during the completion of this work. SAJ and MT is supported by the research deputy of Sharif University of Technology Grant No. G960214.

\bibliography{mybib}

\begin{thebibliography}{41}%
\makeatletter
\providecommand \@ifxundefined [1]{%
 \@ifx{#1\undefined}
}%
\providecommand \@ifnum [1]{%
 \ifnum #1\expandafter \@firstoftwo
 \else \expandafter \@secondoftwo
 \fi
}%
\providecommand \@ifx [1]{%
 \ifx #1\expandafter \@firstoftwo
 \else \expandafter \@secondoftwo
 \fi
}%
\providecommand \natexlab [1]{#1}%
\providecommand \enquote  [1]{``#1''}%
\providecommand \bibnamefont  [1]{#1}%
\providecommand \bibfnamefont [1]{#1}%
\providecommand \citenamefont [1]{#1}%
\providecommand \href@noop [0]{\@secondoftwo}%
\providecommand \href [0]{\begingroup \@sanitize@url \@href}%
\providecommand \@href[1]{\@@startlink{#1}\@@href}%
\providecommand \@@href[1]{\endgroup#1\@@endlink}%
\providecommand \@sanitize@url [0]{\catcode `\\12\catcode `\$12\catcode
  `\&12\catcode `\#12\catcode `\^12\catcode `\_12\catcode `\%12\relax}%
\providecommand \@@startlink[1]{}%
\providecommand \@@endlink[0]{}%
\providecommand \url  [0]{\begingroup\@sanitize@url \@url }%
\providecommand \@url [1]{\endgroup\@href {#1}{\urlprefix }}%
\providecommand \urlprefix  [0]{URL }%
\providecommand \Eprint [0]{\href }%
\providecommand \doibase [0]{http://dx.doi.org/}%
\providecommand \selectlanguage [0]{\@gobble}%
\providecommand \bibinfo  [0]{\@secondoftwo}%
\providecommand \bibfield  [0]{\@secondoftwo}%
\providecommand \translation [1]{[#1]}%
\providecommand \BibitemOpen [0]{}%
\providecommand \bibitemStop [0]{}%
\providecommand \bibitemNoStop [0]{.\EOS\space}%
\providecommand \EOS [0]{\spacefactor3000\relax}%
\providecommand \BibitemShut  [1]{\csname bibitem#1\endcsname}%
\let\auto@bib@innerbib\@empty
\bibitem [{\citenamefont {Wehling}\ \emph {et~al.}(2014)\citenamefont
  {Wehling}, \citenamefont {Black-Schaffer},\ and\ \citenamefont
  {Balatsky}}]{WehlingReview}%
  \BibitemOpen
  \bibfield  {author} {\bibinfo {author} {\bibfnamefont {T.~O.}\ \bibnamefont
  {Wehling}}, \bibinfo {author} {\bibfnamefont {A.~M.}\ \bibnamefont
  {Black-Schaffer}}, \ and\ \bibinfo {author} {\bibfnamefont {A.~V.}\
  \bibnamefont {Balatsky}},\ }\href {\doibase 10.1080/00018732.2014.927109}
  {\bibfield  {journal} {\bibinfo  {journal} {Adv. Phys.}\ }\textbf {\bibinfo
  {volume} {63}},\ \bibinfo {pages} {1} (\bibinfo {year} {2014})}\BibitemShut
  {NoStop}%
\bibitem [{\citenamefont {Katsnelson}(2012)}]{KatsnelsonBook2012}%
  \BibitemOpen
  \bibfield  {author} {\bibinfo {author} {\bibfnamefont {M.~I.}\ \bibnamefont
  {Katsnelson}},\ }\href {\doibase 10.1017/cbo9781139031080} {\emph {\bibinfo
  {title} {Graphene}}}\ (\bibinfo  {publisher} {Cambridge University Press},\
  \bibinfo {year} {2012})\BibitemShut {NoStop}%
\bibitem [{\citenamefont {Crossno}\ \emph {et~al.}(2016)\citenamefont
  {Crossno}, \citenamefont {Shi}, \citenamefont {Wang}, \citenamefont {Liu},
  \citenamefont {Harzheim}, \citenamefont {Lucas}, \citenamefont {Sachdev},
  \citenamefont {Kim}, \citenamefont {Taniguchi}, \citenamefont {Watanabe},
  \citenamefont {Ohki},\ and\ \citenamefont {Fong}}]{Crossno_2016}%
  \BibitemOpen
  \bibfield  {author} {\bibinfo {author} {\bibfnamefont {J.}~\bibnamefont
  {Crossno}}, \bibinfo {author} {\bibfnamefont {J.~K.}\ \bibnamefont {Shi}},
  \bibinfo {author} {\bibfnamefont {K.}~\bibnamefont {Wang}}, \bibinfo {author}
  {\bibfnamefont {X.}~\bibnamefont {Liu}}, \bibinfo {author} {\bibfnamefont
  {A.}~\bibnamefont {Harzheim}}, \bibinfo {author} {\bibfnamefont
  {A.}~\bibnamefont {Lucas}}, \bibinfo {author} {\bibfnamefont
  {S.}~\bibnamefont {Sachdev}}, \bibinfo {author} {\bibfnamefont
  {P.}~\bibnamefont {Kim}}, \bibinfo {author} {\bibfnamefont {T.}~\bibnamefont
  {Taniguchi}}, \bibinfo {author} {\bibfnamefont {K.}~\bibnamefont {Watanabe}},
  \bibinfo {author} {\bibfnamefont {T.~A.}\ \bibnamefont {Ohki}}, \ and\
  \bibinfo {author} {\bibfnamefont {K.~C.}\ \bibnamefont {Fong}},\ }\href
  {\doibase 10.1126/science.aad0343} {\bibfield  {journal} {\bibinfo  {journal}
  {Science}\ }\textbf {\bibinfo {volume} {351}},\ \bibinfo {pages} {1058}
  (\bibinfo {year} {2016})}\BibitemShut {NoStop}%
\bibitem [{\citenamefont {Lucas}\ and\ \citenamefont
  {Fong}(2018)}]{Lucas:2017idv}%
  \BibitemOpen
  \bibfield  {author} {\bibinfo {author} {\bibfnamefont {A.}~\bibnamefont
  {Lucas}}\ and\ \bibinfo {author} {\bibfnamefont {K.~C.}\ \bibnamefont
  {Fong}},\ }\href {\doibase 10.1088/1361-648X/aaa274} {\bibfield  {journal}
  {\bibinfo  {journal} {J. Phys. Condens. Matter}\ }\textbf {\bibinfo {volume}
  {30}},\ \bibinfo {pages} {053001} (\bibinfo {year} {2018})},\ \Eprint
  {http://arxiv.org/abs/1710.08425} {arXiv:1710.08425 [cond-mat.str-el]}
  \BibitemShut {NoStop}%
\bibitem [{\citenamefont {Maldacena}(1998)}]{Maldacena}%
  \BibitemOpen
  \bibfield  {author} {\bibinfo {author} {\bibfnamefont {J.~M.}\ \bibnamefont
  {Maldacena}},\ }\href {\doibase 10.1023/A:1026654312961} {\bibfield
  {journal} {\bibinfo  {journal} {Adv. Theor. Math. Phys.}\ }\textbf {\bibinfo
  {volume} {2}},\ \bibinfo {pages} {231} (\bibinfo {year} {1998})},\ \Eprint
  {http://arxiv.org/abs/hep-th/9711200} {arXiv:hep-th/9711200} \BibitemShut
  {NoStop}%
\bibitem [{\citenamefont {Gubser}\ \emph {et~al.}(1998)\citenamefont {Gubser},
  \citenamefont {Klebanov},\ and\ \citenamefont {Polyakov}}]{GKP}%
  \BibitemOpen
  \bibfield  {author} {\bibinfo {author} {\bibfnamefont {S.~S.}\ \bibnamefont
  {Gubser}}, \bibinfo {author} {\bibfnamefont {I.~R.}\ \bibnamefont
  {Klebanov}}, \ and\ \bibinfo {author} {\bibfnamefont {A.~M.}\ \bibnamefont
  {Polyakov}},\ }\href {\doibase 10.1016/S0370-2693(98)00377-3} {\bibfield
  {journal} {\bibinfo  {journal} {Phys. Lett. B}\ }\textbf {\bibinfo {volume}
  {428}},\ \bibinfo {pages} {105} (\bibinfo {year} {1998})},\ \Eprint
  {http://arxiv.org/abs/hep-th/9802109} {arXiv:hep-th/9802109} \BibitemShut
  {NoStop}%
\bibitem [{\citenamefont {Witten}(1998{\natexlab{a}})}]{Witten}%
  \BibitemOpen
  \bibfield  {author} {\bibinfo {author} {\bibfnamefont {E.}~\bibnamefont
  {Witten}},\ }\href {\doibase 10.4310/ATMP.1998.v2.n3.a3} {\bibfield
  {journal} {\bibinfo  {journal} {Adv. Theor. Math. Phys.}\ }\textbf {\bibinfo
  {volume} {2}},\ \bibinfo {pages} {505} (\bibinfo {year}
  {1998}{\natexlab{a}})},\ \Eprint {http://arxiv.org/abs/hep-th/9803131}
  {arXiv:hep-th/9803131} \BibitemShut {NoStop}%
\bibitem [{\citenamefont {Witten}(1998{\natexlab{b}})}]{Witten:1998zw}%
  \BibitemOpen
  \bibfield  {author} {\bibinfo {author} {\bibfnamefont {E.}~\bibnamefont
  {Witten}},\ }\href {\doibase 10.4310/ATMP.1998.v2.n3.a3} {\bibfield
  {journal} {\bibinfo  {journal} {Adv. Theor. Math. Phys.}\ }\textbf {\bibinfo
  {volume} {2}},\ \bibinfo {pages} {505} (\bibinfo {year}
  {1998}{\natexlab{b}})},\ \Eprint {http://arxiv.org/abs/hep-th/9803131}
  {arXiv:hep-th/9803131} \BibitemShut {NoStop}%
\bibitem [{\citenamefont {Kovtun}\ \emph {et~al.}(2003)\citenamefont {Kovtun},
  \citenamefont {Son},\ and\ \citenamefont {Starinets}}]{KSS-1}%
  \BibitemOpen
  \bibfield  {author} {\bibinfo {author} {\bibfnamefont {P.}~\bibnamefont
  {Kovtun}}, \bibinfo {author} {\bibfnamefont {D.~T.}\ \bibnamefont {Son}}, \
  and\ \bibinfo {author} {\bibfnamefont {A.~O.}\ \bibnamefont {Starinets}},\
  }\href {\doibase 10.1088/1126-6708/2003/10/064} {\bibfield  {journal}
  {\bibinfo  {journal} {JHEP}\ }\textbf {\bibinfo {volume} {10}},\ \bibinfo
  {pages} {064} (\bibinfo {year} {2003})},\ \Eprint
  {http://arxiv.org/abs/hep-th/0309213} {arXiv:hep-th/0309213} \BibitemShut
  {NoStop}%
\bibitem [{\citenamefont {Kovtun}\ \emph {et~al.}(2005)\citenamefont {Kovtun},
  \citenamefont {Son},\ and\ \citenamefont {Starinets}}]{KSS-2}%
  \BibitemOpen
  \bibfield  {author} {\bibinfo {author} {\bibfnamefont {P.}~\bibnamefont
  {Kovtun}}, \bibinfo {author} {\bibfnamefont {D.~T.}\ \bibnamefont {Son}}, \
  and\ \bibinfo {author} {\bibfnamefont {A.~O.}\ \bibnamefont {Starinets}},\
  }\href {\doibase 10.1103/PhysRevLett.94.111601} {\bibfield  {journal}
  {\bibinfo  {journal} {Phys. Rev. Lett.}\ }\textbf {\bibinfo {volume} {94}},\
  \bibinfo {pages} {111601} (\bibinfo {year} {2005})},\ \Eprint
  {http://arxiv.org/abs/hep-th/0405231} {arXiv:hep-th/0405231} \BibitemShut
  {NoStop}%
\bibitem [{\citenamefont {Policastro}\ \emph {et~al.}(2001)\citenamefont
  {Policastro}, \citenamefont {Son},\ and\ \citenamefont {Starinets}}]{PSS}%
  \BibitemOpen
  \bibfield  {author} {\bibinfo {author} {\bibfnamefont {G.}~\bibnamefont
  {Policastro}}, \bibinfo {author} {\bibfnamefont {D.~T.}\ \bibnamefont {Son}},
  \ and\ \bibinfo {author} {\bibfnamefont {A.~O.}\ \bibnamefont {Starinets}},\
  }\href {\doibase 10.1103/PhysRevLett.87.081601} {\bibfield  {journal}
  {\bibinfo  {journal} {Phys. Rev. Lett.}\ }\textbf {\bibinfo {volume} {87}},\
  \bibinfo {pages} {081601} (\bibinfo {year} {2001})},\ \Eprint
  {http://arxiv.org/abs/hep-th/0104066} {arXiv:hep-th/0104066} \BibitemShut
  {NoStop}%
\bibitem [{\citenamefont {Buchel}\ and\ \citenamefont
  {Liu}(2004)}]{Buchel:2003tz}%
  \BibitemOpen
  \bibfield  {author} {\bibinfo {author} {\bibfnamefont {A.}~\bibnamefont
  {Buchel}}\ and\ \bibinfo {author} {\bibfnamefont {J.~T.}\ \bibnamefont
  {Liu}},\ }\href {\doibase 10.1103/PhysRevLett.93.090602} {\bibfield
  {journal} {\bibinfo  {journal} {Phys. Rev. Lett.}\ }\textbf {\bibinfo
  {volume} {93}},\ \bibinfo {pages} {090602} (\bibinfo {year} {2004})},\
  \Eprint {http://arxiv.org/abs/hep-th/0311175} {arXiv:hep-th/0311175}
  \BibitemShut {NoStop}%
\bibitem [{\citenamefont {Shuryak}(2004)}]{Shuryak:2003xe}%
  \BibitemOpen
  \bibfield  {author} {\bibinfo {author} {\bibfnamefont {E.}~\bibnamefont
  {Shuryak}},\ }\href {\doibase 10.1016/j.ppnp.2004.02.025} {\bibfield
  {journal} {\bibinfo  {journal} {Prog. Part. Nucl. Phys.}\ }\textbf {\bibinfo
  {volume} {53}},\ \bibinfo {pages} {273} (\bibinfo {year} {2004})},\ \Eprint
  {http://arxiv.org/abs/hep-ph/0312227} {arXiv:hep-ph/0312227} \BibitemShut
  {NoStop}%
\bibitem [{\citenamefont {Cao}\ \emph {et~al.}(2011)\citenamefont {Cao},
  \citenamefont {Elliott}, \citenamefont {Joseph}, \citenamefont {Wu},
  \citenamefont {Petricka}, \citenamefont {Sch\"afer},\ and\ \citenamefont
  {Thomas}}]{Cao:2010wa}%
  \BibitemOpen
  \bibfield  {author} {\bibinfo {author} {\bibfnamefont {C.}~\bibnamefont
  {Cao}}, \bibinfo {author} {\bibfnamefont {E.}~\bibnamefont {Elliott}},
  \bibinfo {author} {\bibfnamefont {J.}~\bibnamefont {Joseph}}, \bibinfo
  {author} {\bibfnamefont {H.}~\bibnamefont {Wu}}, \bibinfo {author}
  {\bibfnamefont {J.}~\bibnamefont {Petricka}}, \bibinfo {author}
  {\bibfnamefont {T.}~\bibnamefont {Sch\"afer}}, \ and\ \bibinfo {author}
  {\bibfnamefont {J.~E.}\ \bibnamefont {Thomas}},\ }\href {\doibase
  10.1126/science.1195219} {\bibfield  {journal} {\bibinfo  {journal}
  {Science}\ }\textbf {\bibinfo {volume} {331}},\ \bibinfo {pages} {58}
  (\bibinfo {year} {2011})},\ \Eprint {http://arxiv.org/abs/1007.2625}
  {arXiv:1007.2625 [cond-mat.quant-gas]} \BibitemShut {NoStop}%
\bibitem [{\citenamefont {Sch\"afer}\ and\ \citenamefont
  {Teaney}(2009)}]{Schafer:2009dj}%
  \BibitemOpen
  \bibfield  {author} {\bibinfo {author} {\bibfnamefont {T.}~\bibnamefont
  {Sch\"afer}}\ and\ \bibinfo {author} {\bibfnamefont {D.}~\bibnamefont
  {Teaney}},\ }\href {\doibase 10.1088/0034-4885/72/12/126001} {\bibfield
  {journal} {\bibinfo  {journal} {Rept. Prog. Phys.}\ }\textbf {\bibinfo
  {volume} {72}},\ \bibinfo {pages} {126001} (\bibinfo {year} {2009})},\
  \Eprint {http://arxiv.org/abs/0904.3107} {arXiv:0904.3107 [hep-ph]}
  \BibitemShut {NoStop}%
\bibitem [{\citenamefont {Cremonini}(2011)}]{Cremonini:2011iq}%
  \BibitemOpen
  \bibfield  {author} {\bibinfo {author} {\bibfnamefont {S.}~\bibnamefont
  {Cremonini}},\ }\href {\doibase 10.1142/S0217984911027315} {\bibfield
  {journal} {\bibinfo  {journal} {Mod. Phys. Lett. B}\ }\textbf {\bibinfo
  {volume} {25}},\ \bibinfo {pages} {1867} (\bibinfo {year} {2011})},\ \Eprint
  {http://arxiv.org/abs/1108.0677} {arXiv:1108.0677 [hep-th]} \BibitemShut
  {NoStop}%
\bibitem [{\citenamefont {M\"uller}\ \emph {et~al.}(2009)\citenamefont
  {M\"uller}, \citenamefont {Schmalian},\ and\ \citenamefont
  {Fritz}}]{Muller2009}%
  \BibitemOpen
  \bibfield  {author} {\bibinfo {author} {\bibfnamefont {M.}~\bibnamefont
  {M\"uller}}, \bibinfo {author} {\bibfnamefont {J.}~\bibnamefont {Schmalian}},
  \ and\ \bibinfo {author} {\bibfnamefont {L.}~\bibnamefont {Fritz}},\ }\href
  {\doibase 10.1103/PhysRevLett.103.025301} {\bibfield  {journal} {\bibinfo
  {journal} {Phys. Rev. Lett.}\ }\textbf {\bibinfo {volume} {103}},\ \bibinfo
  {pages} {025301} (\bibinfo {year} {2009})}\BibitemShut {NoStop}%
\bibitem [{\citenamefont {Kats}\ and\ \citenamefont
  {Petrov}(2009)}]{Kats:2007mq}%
  \BibitemOpen
  \bibfield  {author} {\bibinfo {author} {\bibfnamefont {Y.}~\bibnamefont
  {Kats}}\ and\ \bibinfo {author} {\bibfnamefont {P.}~\bibnamefont {Petrov}},\
  }\href {\doibase 10.1088/1126-6708/2009/01/044} {\bibfield  {journal}
  {\bibinfo  {journal} {JHEP}\ }\textbf {\bibinfo {volume} {01}},\ \bibinfo
  {pages} {044} (\bibinfo {year} {2009})},\ \Eprint
  {http://arxiv.org/abs/0712.0743} {arXiv:0712.0743 [hep-th]} \BibitemShut
  {NoStop}%
\bibitem [{\citenamefont {Brigante}\ \emph {et~al.}(2008)\citenamefont
  {Brigante}, \citenamefont {Liu}, \citenamefont {Myers}, \citenamefont
  {Shenker},\ and\ \citenamefont {Yaida}}]{Brigante:2007nu}%
  \BibitemOpen
  \bibfield  {author} {\bibinfo {author} {\bibfnamefont {M.}~\bibnamefont
  {Brigante}}, \bibinfo {author} {\bibfnamefont {H.}~\bibnamefont {Liu}},
  \bibinfo {author} {\bibfnamefont {R.~C.}\ \bibnamefont {Myers}}, \bibinfo
  {author} {\bibfnamefont {S.}~\bibnamefont {Shenker}}, \ and\ \bibinfo
  {author} {\bibfnamefont {S.}~\bibnamefont {Yaida}},\ }\href {\doibase
  10.1103/PhysRevD.77.126006} {\bibfield  {journal} {\bibinfo  {journal} {Phys.
  Rev. D}\ }\textbf {\bibinfo {volume} {77}},\ \bibinfo {pages} {126006}
  (\bibinfo {year} {2008})},\ \Eprint {http://arxiv.org/abs/0712.0805}
  {arXiv:0712.0805 [hep-th]} \BibitemShut {NoStop}%
\bibitem [{\citenamefont {Alberte}\ \emph {et~al.}(2016)\citenamefont
  {Alberte}, \citenamefont {Baggioli},\ and\ \citenamefont
  {Pujolas}}]{Alberte:2016xja}%
  \BibitemOpen
  \bibfield  {author} {\bibinfo {author} {\bibfnamefont {L.}~\bibnamefont
  {Alberte}}, \bibinfo {author} {\bibfnamefont {M.}~\bibnamefont {Baggioli}}, \
  and\ \bibinfo {author} {\bibfnamefont {O.}~\bibnamefont {Pujolas}},\ }\href
  {\doibase 10.1007/JHEP07(2016)074} {\bibfield  {journal} {\bibinfo  {journal}
  {JHEP}\ }\textbf {\bibinfo {volume} {07}},\ \bibinfo {pages} {074} (\bibinfo
  {year} {2016})},\ \Eprint {http://arxiv.org/abs/1601.03384} {arXiv:1601.03384
  [hep-th]} \BibitemShut {NoStop}%
\bibitem [{\citenamefont {Rebhan}\ and\ \citenamefont
  {Steineder}(2012)}]{Rebhan:2011vd}%
  \BibitemOpen
  \bibfield  {author} {\bibinfo {author} {\bibfnamefont {A.}~\bibnamefont
  {Rebhan}}\ and\ \bibinfo {author} {\bibfnamefont {D.}~\bibnamefont
  {Steineder}},\ }\href {\doibase 10.1103/PhysRevLett.108.021601} {\bibfield
  {journal} {\bibinfo  {journal} {Phys. Rev. Lett.}\ }\textbf {\bibinfo
  {volume} {108}},\ \bibinfo {pages} {021601} (\bibinfo {year} {2012})},\
  \Eprint {http://arxiv.org/abs/1110.6825} {arXiv:1110.6825 [hep-th]}
  \BibitemShut {NoStop}%
\bibitem [{\citenamefont {Jain}\ \emph
  {et~al.}(2015{\natexlab{a}})\citenamefont {Jain}, \citenamefont {Samanta},\
  and\ \citenamefont {Trivedi}}]{Jain}%
  \BibitemOpen
  \bibfield  {author} {\bibinfo {author} {\bibfnamefont {S.}~\bibnamefont
  {Jain}}, \bibinfo {author} {\bibfnamefont {R.}~\bibnamefont {Samanta}}, \
  and\ \bibinfo {author} {\bibfnamefont {S.~P.}\ \bibnamefont {Trivedi}},\
  }\href {\doibase 10.1007/JHEP10(2015)028} {\bibfield  {journal} {\bibinfo
  {journal} {JHEP}\ }\textbf {\bibinfo {volume} {10}},\ \bibinfo {pages} {028}
  (\bibinfo {year} {2015}{\natexlab{a}})},\ \Eprint
  {http://arxiv.org/abs/1506.01899} {arXiv:1506.01899 [hep-th]} \BibitemShut
  {NoStop}%
\bibitem [{\citenamefont {Jain}\ \emph
  {et~al.}(2015{\natexlab{b}})\citenamefont {Jain}, \citenamefont {Kundu},
  \citenamefont {Sen}, \citenamefont {Sinha},\ and\ \citenamefont
  {Trivedi}}]{Jain2}%
  \BibitemOpen
  \bibfield  {author} {\bibinfo {author} {\bibfnamefont {S.}~\bibnamefont
  {Jain}}, \bibinfo {author} {\bibfnamefont {N.}~\bibnamefont {Kundu}},
  \bibinfo {author} {\bibfnamefont {K.}~\bibnamefont {Sen}}, \bibinfo {author}
  {\bibfnamefont {A.}~\bibnamefont {Sinha}}, \ and\ \bibinfo {author}
  {\bibfnamefont {S.~P.}\ \bibnamefont {Trivedi}},\ }\href {\doibase
  10.1007/JHEP01(2015)005} {\bibfield  {journal} {\bibinfo  {journal} {JHEP}\
  }\textbf {\bibinfo {volume} {01}},\ \bibinfo {pages} {005} (\bibinfo {year}
  {2015}{\natexlab{b}})},\ \Eprint {http://arxiv.org/abs/1406.4874}
  {arXiv:1406.4874 [hep-th]} \BibitemShut {NoStop}%
\bibitem [{\citenamefont {Mamo}(2012)}]{Mamo}%
  \BibitemOpen
  \bibfield  {author} {\bibinfo {author} {\bibfnamefont {K.~A.}\ \bibnamefont
  {Mamo}},\ }\href {\doibase 10.1007/JHEP10(2012)070} {\bibfield  {journal}
  {\bibinfo  {journal} {JHEP}\ }\textbf {\bibinfo {volume} {10}},\ \bibinfo
  {pages} {070} (\bibinfo {year} {2012})},\ \Eprint
  {http://arxiv.org/abs/1205.1797} {arXiv:1205.1797 [hep-th]} \BibitemShut
  {NoStop}%
\bibitem [{\citenamefont {Critelli}\ \emph {et~al.}(2014)\citenamefont
  {Critelli}, \citenamefont {Finazzo}, \citenamefont {Zaniboni},\ and\
  \citenamefont {Noronha}}]{Critelli}%
  \BibitemOpen
  \bibfield  {author} {\bibinfo {author} {\bibfnamefont {R.}~\bibnamefont
  {Critelli}}, \bibinfo {author} {\bibfnamefont {S.~I.}\ \bibnamefont
  {Finazzo}}, \bibinfo {author} {\bibfnamefont {M.}~\bibnamefont {Zaniboni}}, \
  and\ \bibinfo {author} {\bibfnamefont {J.}~\bibnamefont {Noronha}},\ }\href
  {\doibase 10.1103/PhysRevD.90.066006} {\bibfield  {journal} {\bibinfo
  {journal} {Phys. Rev. D}\ }\textbf {\bibinfo {volume} {90}},\ \bibinfo
  {pages} {066006} (\bibinfo {year} {2014})},\ \Eprint
  {http://arxiv.org/abs/1406.6019} {arXiv:1406.6019 [hep-th]} \BibitemShut
  {NoStop}%
\bibitem [{\citenamefont {Erdmenger}\ \emph {et~al.}(2011)\citenamefont
  {Erdmenger}, \citenamefont {Kerner},\ and\ \citenamefont
  {Zeller}}]{Erdmenger}%
  \BibitemOpen
  \bibfield  {author} {\bibinfo {author} {\bibfnamefont {J.}~\bibnamefont
  {Erdmenger}}, \bibinfo {author} {\bibfnamefont {P.}~\bibnamefont {Kerner}}, \
  and\ \bibinfo {author} {\bibfnamefont {H.}~\bibnamefont {Zeller}},\ }\href
  {\doibase 10.1016/j.physletb.2011.04.009} {\bibfield  {journal} {\bibinfo
  {journal} {Phys. Lett. B}\ }\textbf {\bibinfo {volume} {699}},\ \bibinfo
  {pages} {301} (\bibinfo {year} {2011})},\ \Eprint
  {http://arxiv.org/abs/1011.5912} {arXiv:1011.5912 [hep-th]} \BibitemShut
  {NoStop}%
\bibitem [{\citenamefont {Katayama}\ \emph {et~al.}(2006)\citenamefont
  {Katayama}, \citenamefont {Kobayashi},\ and\ \citenamefont
  {Suzumura}}]{Katayama2006}%
  \BibitemOpen
  \bibfield  {author} {\bibinfo {author} {\bibfnamefont {S.}~\bibnamefont
  {Katayama}}, \bibinfo {author} {\bibfnamefont {A.}~\bibnamefont {Kobayashi}},
  \ and\ \bibinfo {author} {\bibfnamefont {Y.}~\bibnamefont {Suzumura}},\
  }\href {\doibase 10.1143/jpsj.75.054705} {\bibfield  {journal} {\bibinfo
  {journal} {Journal of the Physical Society of Japan}\ }\textbf {\bibinfo
  {volume} {75}},\ \bibinfo {pages} {054705} (\bibinfo {year}
  {2006})}\BibitemShut {NoStop}%
\bibitem [{\citenamefont {Zhou}\ \emph {et~al.}(2014)\citenamefont {Zhou},
  \citenamefont {Dong}, \citenamefont {Oganov}, \citenamefont {Zhu},
  \citenamefont {Tian},\ and\ \citenamefont {Wang}}]{Zhou2014}%
  \BibitemOpen
  \bibfield  {author} {\bibinfo {author} {\bibfnamefont {X.-F.}\ \bibnamefont
  {Zhou}}, \bibinfo {author} {\bibfnamefont {X.}~\bibnamefont {Dong}}, \bibinfo
  {author} {\bibfnamefont {A.~R.}\ \bibnamefont {Oganov}}, \bibinfo {author}
  {\bibfnamefont {Q.}~\bibnamefont {Zhu}}, \bibinfo {author} {\bibfnamefont
  {Y.}~\bibnamefont {Tian}}, \ and\ \bibinfo {author} {\bibfnamefont {H.-T.}\
  \bibnamefont {Wang}},\ }\href {\doibase 10.1103/PhysRevLett.112.085502}
  {\bibfield  {journal} {\bibinfo  {journal} {Phys. Rev. Lett.}\ }\textbf
  {\bibinfo {volume} {112}},\ \bibinfo {pages} {085502} (\bibinfo {year}
  {2014})}\BibitemShut {NoStop}%
\bibitem [{\citenamefont {Goerbig}\ \emph {et~al.}(2008)\citenamefont
  {Goerbig}, \citenamefont {Fuchs}, \citenamefont {Montambaux},\ and\
  \citenamefont {Pi\'echon}}]{Goerbig2008}%
  \BibitemOpen
  \bibfield  {author} {\bibinfo {author} {\bibfnamefont {M.~O.}\ \bibnamefont
  {Goerbig}}, \bibinfo {author} {\bibfnamefont {J.-N.}\ \bibnamefont {Fuchs}},
  \bibinfo {author} {\bibfnamefont {G.}~\bibnamefont {Montambaux}}, \ and\
  \bibinfo {author} {\bibfnamefont {F.}~\bibnamefont {Pi\'echon}},\ }\href
  {\doibase 10.1103/PhysRevB.78.045415} {\bibfield  {journal} {\bibinfo
  {journal} {Phys. Rev. B}\ }\textbf {\bibinfo {volume} {78}},\ \bibinfo
  {pages} {045415} (\bibinfo {year} {2008})}\BibitemShut {NoStop}%
\bibitem [{\citenamefont {Volovik}(2016)}]{Volovik2016}%
  \BibitemOpen
  \bibfield  {author} {\bibinfo {author} {\bibfnamefont {G.~E.}\ \bibnamefont
  {Volovik}},\ }\href {\doibase 10.1134/s0021364016210050} {\bibfield
  {journal} {\bibinfo  {journal} {{JETP} Letters}\ }\textbf {\bibinfo {volume}
  {104}},\ \bibinfo {pages} {645} (\bibinfo {year} {2016})}\BibitemShut
  {NoStop}%
\bibitem [{\citenamefont {Farajollahpour}\ \emph {et~al.}(2019)\citenamefont
  {Farajollahpour}, \citenamefont {Faraei},\ and\ \citenamefont
  {Jafari}}]{Tohid2019Spacetime}%
  \BibitemOpen
  \bibfield  {author} {\bibinfo {author} {\bibfnamefont {T.}~\bibnamefont
  {Farajollahpour}}, \bibinfo {author} {\bibfnamefont {Z.}~\bibnamefont
  {Faraei}}, \ and\ \bibinfo {author} {\bibfnamefont {S.~A.}\ \bibnamefont
  {Jafari}},\ }\href {\doibase 10.1103/PhysRevB.99.235150} {\bibfield
  {journal} {\bibinfo  {journal} {Phys. Rev. B}\ }\textbf {\bibinfo {volume}
  {99}},\ \bibinfo {pages} {235150} (\bibinfo {year} {2019})}\BibitemShut
  {NoStop}%
\bibitem [{\citenamefont {Jafari}(2019)}]{Jafari2019}%
  \BibitemOpen
  \bibfield  {author} {\bibinfo {author} {\bibfnamefont {S.~A.}\ \bibnamefont
  {Jafari}},\ }\href {\doibase 10.1103/physrevb.100.045144} {\bibfield
  {journal} {\bibinfo  {journal} {Physical Review B}\ }\textbf {\bibinfo
  {volume} {100}} (\bibinfo {year} {2019}),\
  10.1103/physrevb.100.045144}\BibitemShut {NoStop}%
\bibitem [{\citenamefont {Farajollahpour}\ and\ \citenamefont
  {Jafari}(2020)}]{Farajollahpour_2020}%
  \BibitemOpen
  \bibfield  {author} {\bibinfo {author} {\bibfnamefont {T.}~\bibnamefont
  {Farajollahpour}}\ and\ \bibinfo {author} {\bibfnamefont {S.~A.}\
  \bibnamefont {Jafari}},\ }\href {\doibase 10.1103/physrevresearch.2.023410}
  {\bibfield  {journal} {\bibinfo  {journal} {Physical Review Research}\
  }\textbf {\bibinfo {volume} {2}} (\bibinfo {year} {2020}),\
  10.1103/physrevresearch.2.023410}\BibitemShut {NoStop}%
\bibitem [{\citenamefont {Weststr\"om}\ and\ \citenamefont
  {Ojanen}(2017)}]{OjanenPRX}%
  \BibitemOpen
  \bibfield  {author} {\bibinfo {author} {\bibfnamefont {A.}~\bibnamefont
  {Weststr\"om}}\ and\ \bibinfo {author} {\bibfnamefont {T.}~\bibnamefont
  {Ojanen}},\ }\href {\doibase 10.1103/PhysRevX.7.041026} {\bibfield  {journal}
  {\bibinfo  {journal} {Phys. Rev. X}\ }\textbf {\bibinfo {volume} {7}},\
  \bibinfo {pages} {041026} (\bibinfo {year} {2017})}\BibitemShut {NoStop}%
\bibitem [{\citenamefont {Moradpouri}\ \emph {et~al.}(2020)\citenamefont
  {Moradpouri}, \citenamefont {Torabian},\ and\ \citenamefont
  {Jafari}}]{Moradpouri:2020wwa}%
  \BibitemOpen
  \bibfield  {author} {\bibinfo {author} {\bibfnamefont {A.}~\bibnamefont
  {Moradpouri}}, \bibinfo {author} {\bibfnamefont {M.}~\bibnamefont
  {Torabian}}, \ and\ \bibinfo {author} {\bibfnamefont {S.~A.}\ \bibnamefont
  {Jafari}},\ }\href@noop {} {\  (\bibinfo {year} {2020})},\ \Eprint
  {http://arxiv.org/abs/2007.03276} {arXiv:2007.03276 [cond-mat.mes-hall]}
  \BibitemShut {NoStop}%
\bibitem [{\citenamefont {Hubeny}\ \emph {et~al.}(2012)\citenamefont {Hubeny},
  \citenamefont {Minwalla},\ and\ \citenamefont {Rangamani}}]{Hubeny}%
  \BibitemOpen
  \bibfield  {author} {\bibinfo {author} {\bibfnamefont {V.~E.}\ \bibnamefont
  {Hubeny}}, \bibinfo {author} {\bibfnamefont {S.}~\bibnamefont {Minwalla}}, \
  and\ \bibinfo {author} {\bibfnamefont {M.}~\bibnamefont {Rangamani}},\ }in\
  \href@noop {} {\emph {\bibinfo {booktitle} {{Theoretical Advanced Study
  Institute in Elementary Particle Physics}: {String theory and its
  Applications: From meV to the Planck Scale}}}}\ (\bibinfo {year} {2012})\
  pp.\ \bibinfo {pages} {348--383},\ \Eprint {http://arxiv.org/abs/1107.5780}
  {arXiv:1107.5780 [hep-th]} \BibitemShut {NoStop}%
\bibitem [{\citenamefont {Gibbons}\ and\ \citenamefont
  {Hawking}(1977)}]{Hawking}%
  \BibitemOpen
  \bibfield  {author} {\bibinfo {author} {\bibfnamefont {G.~W.}\ \bibnamefont
  {Gibbons}}\ and\ \bibinfo {author} {\bibfnamefont {S.~W.}\ \bibnamefont
  {Hawking}},\ }\href {\doibase 10.1103/PhysRevD.15.2738} {\bibfield  {journal}
  {\bibinfo  {journal} {Phys. Rev. D}\ }\textbf {\bibinfo {volume} {15}},\
  \bibinfo {pages} {2738} (\bibinfo {year} {1977})}\BibitemShut {NoStop}%
\bibitem [{\citenamefont {de~Haro}\ \emph {et~al.}(2001)\citenamefont
  {de~Haro}, \citenamefont {Solodukhin},\ and\ \citenamefont
  {Skenderis}}]{deHaro}%
  \BibitemOpen
  \bibfield  {author} {\bibinfo {author} {\bibfnamefont {S.}~\bibnamefont
  {de~Haro}}, \bibinfo {author} {\bibfnamefont {S.~N.}\ \bibnamefont
  {Solodukhin}}, \ and\ \bibinfo {author} {\bibfnamefont {K.}~\bibnamefont
  {Skenderis}},\ }\href {\doibase 10.1007/s002200100381} {\bibfield  {journal}
  {\bibinfo  {journal} {Commun. Math. Phys.}\ }\textbf {\bibinfo {volume}
  {217}},\ \bibinfo {pages} {595} (\bibinfo {year} {2001})},\ \Eprint
  {http://arxiv.org/abs/hep-th/0002230} {arXiv:hep-th/0002230} \BibitemShut
  {NoStop}%
\bibitem [{\citenamefont {Balasubramanian}\ and\ \citenamefont
  {Kraus}(1999)}]{Balasubramanian}%
  \BibitemOpen
  \bibfield  {author} {\bibinfo {author} {\bibfnamefont {V.}~\bibnamefont
  {Balasubramanian}}\ and\ \bibinfo {author} {\bibfnamefont {P.}~\bibnamefont
  {Kraus}},\ }\href {\doibase 10.1007/s002200050764} {\bibfield  {journal}
  {\bibinfo  {journal} {Commun. Math. Phys.}\ }\textbf {\bibinfo {volume}
  {208}},\ \bibinfo {pages} {413} (\bibinfo {year} {1999})},\ \Eprint
  {http://arxiv.org/abs/hep-th/9902121} {arXiv:hep-th/9902121} \BibitemShut
  {NoStop}%
\bibitem [{\citenamefont {Ammon}\ and\ \citenamefont
  {Erdmenger}(2015)}]{ammon2015gauge}%
  \BibitemOpen
  \bibfield  {author} {\bibinfo {author} {\bibfnamefont {M.}~\bibnamefont
  {Ammon}}\ and\ \bibinfo {author} {\bibfnamefont {J.}~\bibnamefont
  {Erdmenger}},\ }\href {https://books.google.com/books?id=OG0GBwAAQBAJ} {\emph
  {\bibinfo {title} {Gauge/Gravity Duality}}}\ (\bibinfo  {publisher}
  {Cambridge University Press},\ \bibinfo {year} {2015})\BibitemShut {NoStop}%
\bibitem [{\citenamefont {Son}\ and\ \citenamefont
  {Starinets}(2002)}]{Son_2002}%
  \BibitemOpen
  \bibfield  {author} {\bibinfo {author} {\bibfnamefont {D.~T.}\ \bibnamefont
  {Son}}\ and\ \bibinfo {author} {\bibfnamefont {A.~O.}\ \bibnamefont
  {Starinets}},\ }\href {\doibase 10.1088/1126-6708/2002/09/042} {\bibfield
  {journal} {\bibinfo  {journal} {Journal of High Energy Physics}\ }\textbf
  {\bibinfo {volume} {2002}},\ \bibinfo {pages} {042} (\bibinfo {year}
  {2002})}\BibitemShut {NoStop}%
\end{thebibliography}%

\end{document}